\documentclass[11pt]{amsart}
\usepackage{geometry}                
\usepackage{graphicx}
\usepackage{amssymb}
\usepackage{epstopdf}
\title{An exact solution to the partition function of the finite-size Ising Model}
\author{Rong Qiang Wei}
\address{College of Earth and Planetary Sciences, University of Chinese Academy of Sciences, Beijing, PRC, 100049}
\email{wrq1973@ucas.ac.cn}
\date{}

\begin{document}
\maketitle

\begin{abstract}
  There is no an exact solution to three-dimensional (3D)  finite-size Ising model (referred to as the Ising model hereafter for simplicity) and even two-dimensional (2D) Ising model with non-zero external field to our knowledge. Here by using an elementary but rigorous method, we obtain an exact solution to the partition function of the Ising model with $N$ lattice sites. It is a sum of $2^N$ exponential functions and holds for $D$-dimensional ($D=1,2,3,...$) Ising model with or without the external field. This solution provides a new insight into the problem of the Ising model and the related difficulties, and new understanding of the classic exact solutions for one-dimensional (1D) (Kramers and Wannier, 1941) or 2D Ising model (Onsager, 1944). With this solution, the specific heat and magnetisation of a simple 3D Ising model are calculated, which are consistent with the results from experiments and/or numerical simulations. Furthermore, the solution here and the related approaches, can also be available to other models like the percolation and/or the Potts model.
\end{abstract}

{\hspace{2.2em}\small Keywords:}

{\hspace{2.2em}\tiny Ising model Exact solution Hubbard-Stratonovich transformation Gaussian integral}

\section{Introduction}

The Ising model is a famous statistic mechanical model for the structure of the ferromagnetic substance. This model posits that each lattice site in a material has associated with it a binary magnetic polarity (or "spin"), and the spin may flip to reduce the energy of the system.  Studies on such a system have provided insights into ferromagnetism and phase transitions (eg., Huang, 1987).  These statistic mechanical studies depend on the evaluation of the partition function of the Ising model.

The 1D Ising model is composed of a chain of $N$ spins, in which each spin interacts only with its two nearest neighbors and with an external magnetic field $H$. Usually the periodic boundary condition is imposed. Kramers and Wannier (1941) obtained the partition function $Q(H,T)$ of this model by a method of transfer-matrix as the following,

\begin{equation}\label{KW}
Q(H,T)=\lambda_+^N+\lambda_-^N
\end{equation}
where $$\lambda_{\pm}=\exp(\beta\epsilon)\left[\cosh (\beta H)\pm\sqrt{\sinh ^2(\beta H)+\exp(-4\beta\epsilon)}\right]$$and $\beta=\frac{1}{kT}$. $\epsilon$ is the interaction energy, $k$ Boltzmann constant, $T$  the temperature in K. 

Using the transfer-matrix method, the partition function for the 2D Ising model ($N$ spins; nearest-neighbor coupling; periodic boundary condition) was evaluated exactly by Onsager (1944) for the case of vanishing external magnetic field ($H=0$). This partition function $Q(0,T)$ is, 

\begin{equation}\label{Onsager}
Q(0,T)=(2\cosh 2\beta\epsilon)^N\cdot \exp\left[\frac{N}{2\pi}\int_0^\pi {\rm{d}}\phi\log\frac{1}{2}(1+\sqrt{1-\kappa^2\sin^2\phi})\right]
\end{equation}
where $\phi=\beta\epsilon$, and $$\kappa=\frac{2}{\cosh 2\phi \coth 2\phi}$$

However, the transfer-matrix method is complicated for the 3D Ising model and even 2D Ising models with non-zero external field ($H\neq 0$), and there is still no exact solutions for these problems to our knowledge.  It seems difficult in a short term to find both elegant and exact partition functions like Eq. (\ref{KW}) or (\ref{Onsager}) if no new (mathematical) approaches are introduced. Recently, Kocharovsky and Kocharovsky (2015) found the consistency equations for the main steps in the analysis of the 3D Ising model, towards which an exact solution to the 3D model can be obtained possibly. These equations are based on a rigorous theory of magnetic phase transitions in a mesoscopic lattice of spins which described as the constrained spin bosons in a Holstein-Primakoff representation.

Here we use an elementary method to yield the partition function of the $D$-dimensional ($D=1,2,3,...$) Ising model. Although this solution looks formidable and unintuitive, it is exact and simple. Based on this solution, some thermodynamic functions of a simple 3D Ising model are calculated and discussed. Simultaneously, we hope that the solution and the related method may find other useful application, for example, to the percolation and/or Potts model.

\section{An exact solution to the Ising Model}
\subsection{$H=0$}

The nearest-neighbor Ising model without an external magnetic field in $D$-dimensions is defined in terms of the following Hamiltonian (eg., Huang, 1987),

\begin{equation}\label{eq1}
\mathcal{H} =  -\frac{1}{2}\sum\limits_{i,j = 1}^N {{J_{ij}}} {s_i}{s_j}
\end{equation}
where, $i$ and $j$ are the sites $\bf{r}_i$ and $\bf{r}_j$ of a D-dimensional hyper cubic lattice with $N$ sites, respectively. $s_i=\pm 1$ are the two possible states of the $z$-components of spins localized at the lattice sites. $J_{ij}$ denotes the exchange interaction between spins localized at $\bf{r}_i$ and $\bf{r}_j$, 

\begin{equation}\label{eq2}
{J_{ij}} = \left\{ {\begin{array}{ll}
\epsilon&{{\rm{if \hspace{0.5em}}}i{\rm{\hspace{0.5em}and\hspace{0.5em}}}j{\rm{\hspace{0.5em} are\hspace{0.2em} the \hspace{0.2em}nearest\hspace{0.2em} neighbors}}}\\
0&{{\rm{otherwise}}}
\end{array}} \right.
\end{equation}

Calculations in the later section show that the matrix $\bf{J}$ may be singular.  To avoid this singular we redefine $\bf{J}$  as the following,

 \begin{equation}\label{eq2_1}
{J_{ij}} = \left\{ {\begin{array}{ll}
\epsilon&{{\rm{if \hspace{0.5em}}}i{\rm{\hspace{0.5em}and\hspace{0.5em}}}j{\rm{\hspace{0.5em} are\hspace{0.2em} the \hspace{0.2em}nearest\hspace{0.2em} neighbors}}}\\
a\epsilon & i=j \hspace{0.2em} {\rm{and}} \hspace{0.2em} i\leq \frac{N}{2}, j\leq \frac{N}{2}\\
-a\epsilon & i=j \hspace{0.2em} {\rm{and}} \hspace{0.2em} i > \frac{N}{2}, j > \frac{N}{2}\\ \\
0&{{\rm{otherwise}}}
\end{array}} \right.
\end{equation}
where $a$ is a real number, and here $a=1$ for simplicity.

It should be pointed out that: (1) Other re-definition of $\bf{J}$ is possible, only if the condition that the trace of $\bf{J}$ is zero holds, and the singularity of the original matrix $\bf{J}$ is eliminated. If so, the Hamiltonian (Eq. (\ref{eq1})) is invariant; (2) Here we assume $N$ is even only for simplicity.  

The partition function for this Ising model is,

\begin{equation}\label{eq3}
{\rm Q} = \sum\limits_{\{ {s_i} =  \pm 1\} } {\exp (-\beta \mathcal{H})}= \sum\limits_{\{ {s_i} =  \pm 1\} } {\exp (\frac{1}{2}\sum\limits_{ij} {{K_{ij}}} {s_i}{s_j})}
\end{equation}
where $K_{ij}=\beta J_{ij}$.

By using the following Hubbard-Stratonovich transformation (Stratonovich, 1958; Hubbard, 1959),

\begin{equation}\label{eq4}
\begin{array}{ll}
{\exp (\frac{1}{2}\sum\limits_{ij} {{K_{ij}}} {s_i}{s_j})}\\
{ = {{\left[ {\frac{{\det K}}{{{{(2\pi )}^N}}}} \right]}^{1/2}}\int_{ - \infty }^{ + \infty }  \ldots  \int_{ - \infty }^{ + \infty } {\prod\limits_{k = 1}^N {d{\phi _k}\exp \left( { - \frac{1}{2}\sum\limits_{ij} {{\phi _i}{K_{ij}}{\phi _j} + \sum\limits_{ij} {{s_i}{K_{ij}}{\phi _j}} } } \right)} } }
\end{array}
\end{equation}
and the following relationship,

\begin{equation}\label{eq5}
\begin{array}{ll}
\sum\limits_{\{ {s_i} =  \pm 1\} } \exp \left( {\sum\limits_{ij} {{s_i}{K_{ij}}{\phi _j}} } \right) &= \sum\limits_{\{ {s_i} =  \pm 1\} } {\prod\limits_i {\exp \left( {{s_i}\sum\limits_j {{K_{ij}}{\phi _j}} } \right)} }  \\
&{ = \prod\limits_i {\left[ {\exp \left( {\sum\limits_j {{K_{ij}}{\phi _j}} } \right) + \exp \left( { - \sum\limits_j {{K_{ij}}{\phi _j}} } \right)} \right]} }
\end{array}
\end{equation}

We thereby obtain the partition function for the Ising model as,

\begin{equation}\label{eq6}
\begin{array}{ll}
{\rm Q} &= {{\left[ {\frac{{\det K}}{{{{(2\pi )}^N}}}} \right]}^{1/2}}\int_{ - \infty }^{ + \infty }  \ldots  \int_{ - \infty }^{ + \infty } \prod\limits_{k = 1}^N d{\phi _k}\exp \left( { - \frac{1}{2}\sum\limits_{ij} {{\phi _i}{K_{ij}}{\phi _j}}}  \right) \\
&\ \ \ \times\prod\limits_i {\left[ {\exp \left( {\sum\limits_j {{K_{ij}}{\phi _j}} } \right) + \exp \left( { - \sum\limits_j {{K_{ij}}{\phi _j}} } \right)} \right]} 
\end{array}
\end{equation}
in which the degrees of freedom on the lattice sites expressed in terms of the continuous quantities $\phi_i$. 

Further, we expand $\prod\limits_i {\left[ {\exp \left( {\sum\limits_j {{K_{ij}}{\phi _j}} } \right) + \exp \left( { - \sum\limits_j {{K_{ij}}{\phi _j}} } \right)} \right]}$ as the following,

\begin{equation}\label{eq7}
\begin{array}{ll}
{\prod\limits_i {\left[ {\exp \left( {\sum\limits_j {{K_{ij}}{\phi _j}} } \right) + \exp \left( { - \sum\limits_j {{K_{ij}}{\phi _j}} } \right)} \right]} }\\
{ = \prod\limits_i {\exp \left( {\sum\limits_j {{K_{ij}}{\phi _j}} } \right)\left[ {1 + \exp \left( { - 2\sum\limits_j {{K_{ij}}{\phi _j}} } \right)} \right]} }\\
 = \exp \left( {\sum\limits_i {\sum\limits_j {{K_{ij}}{\phi _j}} } } \right) \cdot \left\{ {1 + \sum\limits_{\alpha  = 1}^N {\exp \left[ { - 2\sum\limits_j {{K_{\alpha j}}{\phi _j}} } \right] + } } \right.  \\
 \sum\limits_{\begin{array}{cc}
{\tiny\alpha ,\beta  = 1}\\
{\alpha  < \beta }
\end{array}}^N {\exp \left[ { - 2\sum\limits_j {({K_{\alpha j}} + {K_{\beta j}}){\phi _j}} } \right]}+  \\
\sum\limits_{\begin{array}{cc}
{\alpha ,\beta ,\gamma  = 1}\\
{\alpha  < \beta  < \gamma }
\end{array}}^N {\exp \left[ { - 2\sum\limits_j {({K_{\alpha j}} + {K_{\beta j}} + {K_{\gamma j}}){\phi _j}} } \right] + \ldots\ldots} \left. \begin{array}{cc}
\ & \  \\
\ & \  \\
\ & \ 
\end{array}
\right\}  
\end{array}
\end{equation}

Let ${\sum\limits_i {\sum\limits_j {{K_{ij}}{\phi _j}} } }=\sum\limits_j{K_{0j}}\phi_j$,
 
\begin{equation}\label{eq8}
\begin{array}{ll}
{\prod\limits_i {\left[ {\exp \left( {\sum\limits_j {{K_{ij}}{\phi _j}} } \right) + \exp \left( { - \sum\limits_j {{K_{ij}}{\phi _j}} } \right)} \right]} }\\
  = {\exp(\sum\limits_jK_{0j}\phi_j) + \sum\limits_{\alpha  = 1}^N {\exp \left[ { \sum\limits_j {(K_{0j}-{2K_{\alpha j}}){\phi _j}} } \right] + } } \\
\sum\limits_{\begin{array}{cc}
{\tiny\alpha ,\beta  = 1}\\
{\alpha  < \beta }
\end{array}}^N \exp \left\{ { \sum\limits_j {[K_{0j}-2({K_{\alpha j}} + {K_{\beta j}})]{\phi _j}} } \right\}+\\
\sum\limits_{\begin{array}{cc}
{\alpha ,\beta ,\gamma  = 1}\\
{\alpha  < \beta  < \gamma }
\end{array}}^N {\exp \left\{ { \sum\limits_j {[K_{0j}-2({K_{\alpha j}} + {K_{\beta j}} + {K_{\gamma j}})]{\phi _j}} } \right\} + }\\
\ \\
 \ldots \\
 \ \\
+\exp(\sum\limits_j -K_{0j}\phi_j)
 \end{array}
\end{equation}

Therefore, the partition function now is, 

\begin{equation}\label{eq9}
\begin{array}{ll}
{\rm Q}=& {{\left[ {\frac{{\det K}}{{{{(2\pi )}^N}}}} \right]}^{1/2}}\left\{\int_{ - \infty }^{ + \infty }  \ldots  \int_{ - \infty }^{ + \infty } \prod\limits_{k = 1}^N d{\phi _k}\exp \left( { - \frac{1}{2}\sum\limits_{ij} {{\phi _i}{K_{ij}}{\phi _j}} } +\sum\limits_jK_{0j}\phi_j\right) +\right. \\
&\sum\limits_{\alpha  = 1}^N \int_{ - \infty }^{ + \infty }  \ldots  \int_{ - \infty }^{ + \infty } \prod\limits_{k = 1}^N d{\phi _k}\exp \left[ { - \frac{1}{2}\sum\limits_{ij} {{\phi _i}{K_{ij}}{\phi _j}} } +\sum\limits_j(K_{0j}-2{K_{\alpha j}})\phi_j\right] +\\
&\sum\limits_{\begin{array}{cc}
{\tiny\alpha ,\beta  = 1}\\
{\alpha  < \beta }
\end{array}}^N \int_{ - \infty }^{ + \infty }  \ldots  \int_{ - \infty }^{ + \infty } \prod\limits_{k = 1}^N d{\phi _k}\exp \left[ { - \frac{1}{2}\sum\limits_{ij} {{\phi _i}{K_{ij}}{\phi _j}} } +\sum\limits_j(K_{0j}-2{K_{\alpha j}}-2{K_{\beta j}})\phi_j\right] +\\
&\sum\limits_{\begin{array}{cc}
{\alpha ,\beta ,\gamma  = 1}\\
{\alpha  < \beta  < \gamma }
\end{array}}^N \int_{ - \infty }^{ + \infty }  \ldots  \int_{ - \infty }^{ + \infty } \prod\limits_{k = 1}^N d{\phi _k}\exp \left[ { - \frac{1}{2}\sum\limits_{ij} {{\phi _i}{K_{ij}}{\phi _j}} } +\sum\limits_j(K_{0j}-2{K_{\alpha j}}-2{K_{\beta j}}-2{K_{\gamma j}})\phi_j\right] +
\ldots \\
&\left. +\int_{ - \infty }^{ + \infty }  \ldots  \int_{ - \infty }^{ + \infty } \prod\limits_{k = 1}^N d{\phi _k}\exp \left( { - \frac{1}{2}\sum\limits_{ij} {{\phi _i}{K_{ij}}{\phi _j}} } +\sum\limits_j -(K_{0j})\phi_j\right)\right\}
 \end{array}
\end{equation}

The N-dimensional Gaussian integrals reads,

\begin{equation}\label{eq10}
\int_{ - \infty }^\infty   \ldots  \int_{ - \infty }^\infty  {\prod\limits_{k = 1} {d{\phi _k}} \exp \left( { - \frac{1}{2}\sum\limits_{ij} {{\phi _i}{K_{ij}}{\phi _j} + \sum\limits_j {{K_{0j}}{\phi _j}} } } \right)}  = {\left[ {\frac{{{{(2\pi )}^N}}}{{\det K}}} \right]^{1/2}}\exp \left( {\frac{1}{2}{\bf{K}}_0{{\bf{K}}^{ - 1}}{{\bf{K}}_0^T}} \right)
\end{equation}
 
Using Eq. (\ref{eq10}), then the partition function now is, 

\begin{equation}\label{eq11}
\begin{array}{ll}
{\rm Q} = & \exp \left( {\frac{1}{2}{\bf{K}}_0{{\bf{K}}^{ - 1}}{{\bf{K}}_0^T}} \right) +
\sum\limits_{\alpha  = 1}^N \exp \left[ {\frac{1}{2}({\bf{K}}_0-2{\bf{K_\alpha}}){{\bf{K}}^{ - 1}}{({\bf{K}}_0-2{\bf{K_\alpha}})^T}} \right]+\\
& \sum\limits_{\begin{array}{cc}
{\tiny\alpha ,\beta  = 1}\\
{\alpha  < \beta }
\end{array}}^N \exp \left[ {\frac{1}{2}({\bf{K}}_0-2{\bf{K_\alpha}}-2{\bf{K_\beta}}){{\bf{K}}^{ - 1}}{({\bf{K}}_0-2{\bf{K_\alpha}}-2{\bf{K_\beta}})^T}} \right]+\\
& \sum\limits_{\begin{array}{cc}
{\alpha ,\beta ,\gamma  = 1}\\
{\alpha  < \beta  < \gamma }
\end{array}}^N \exp \left[ {\frac{1}{2}({\bf{K}}_0-2{\bf{K_\alpha}}-2{\bf{K_\beta}}-2{\bf{K_\gamma}}){{\bf{K}}^{ - 1}}{({\bf{K}}_0-2{\bf{K_\alpha}}-2{\bf{K_\beta}}-2{\bf{K_\gamma}})^T}} \right]+
\ldots \\
& + \exp \left[ {\frac{1}{2}(-{\bf{K}}_0){{\bf{K}}^{ - 1}}{(-{\bf{K}}_0)^T}} \right]
 \end{array}
\end{equation}

\subsection{$H\neq 0$}

The nearest-neighbor Ising model with an external magnetic field in D dimensions is defined in terms of the following Hamiltonian (eg., Huang, 1987),

\begin{equation}\label{eq17}
\mathcal{H} =  -\frac{1}{2}\sum\limits_{i,j = 1}^N {{J_{ij}}} {s_i}{s_j}-h\sum\limits_{i= 1}^N s_i
\end{equation}

where $h=\mu H$ is the Zeeman energy associated with an external magnetic field in the $z$-direction, $\mu$ the permeability.

The partition function for this Ising model is,

\begin{equation}\label{eq18}
{\rm Q} = \sum\limits_{\{ {s_i} =  \pm 1\} } {\exp (-\beta \mathcal{H})}= \sum\limits_{\{ {s_i} =  \pm 1\} } {\exp (\frac{1}{2}\sum\limits_{ij} {{K_{ij}}} {s_i}{s_j}+\beta h \sum\limits_{i= 1}^N s_i)}
\end{equation}

By using the following Hubbard-Stratonovich transformation (Eq. (\ref{eq4})), we can obtain,

\begin{equation}\label{eq19}
\begin{array}{ll}
{\rm Q} & =  \sum\limits_{\{ {s_i} =  \pm 1\} }{{\left[ {\frac{{\det K}}{{{{(2\pi )}^N}}}} \right]}^{1/2}}\int_{ - \infty }^{ + \infty }  \ldots  \int_{ - \infty }^{ + \infty } {\prod\limits_{k = 1}^N {d{\phi _k}\exp \left[ { - \frac{1}{2}\sum\limits_{ij} {{\phi _i}{K_{ij}}{\phi _j} + \sum\limits_{ij} {{s_i}(K_{ij}{\phi _j}+\beta h)} } } \right]} } \\
& = {{\left[ {\frac{{\det K}}{{{{(2\pi )}^N}}}} \right]}^{1/2}}\int_{ - \infty }^{ + \infty }  \ldots  \int_{ - \infty }^{ + \infty } \prod\limits_{k = 1}^N d{\phi _k}\exp \left( { - \frac{1}{2}\sum\limits_{ij} {{\phi _i}{K_{ij}}{\phi _j}} } \right) \\
&\ \ \ \times\prod\limits_i {\left[ {\exp \left(\beta h+{\sum\limits_j {{K_{ij}}{\phi _j}} } \right) + \exp \left( { -\beta h- \sum\limits_j {{K_{ij}}{\phi _j}} } \right)} \right]} 
\end{array}
\end{equation}

Similar to the above,

\begin{equation}\label{eq20}
\begin{array}{ll}
{\rm Q}  = & {{\left[ {\frac{{\det K}}{{{{(2\pi )}^N}}}} \right]}^{1/2}}\left\{ \int_{ - \infty }^{ + \infty }  \ldots  \int_{ - \infty }^{ + \infty } \prod\limits_{k = 1}^N d{\phi _k}\exp \left( { - \frac{1}{2}\sum\limits_{ij} {{\phi _i}{K_{ij}}{\phi _j}} } +\sum\limits_j K_{0j}\phi_j +N \beta h \right) + \right. \\
& \sum\limits_{\alpha  = 1}^N \int_{ - \infty }^{ + \infty }  \ldots  \int_{ - \infty }^{ + \infty } \prod\limits_{k = 1}^N d{\phi _k}\exp \left[ { - \frac{1}{2}\sum\limits_{ij} {{\phi _i}{K_{ij}}{\phi _j}} } +\sum\limits_j(K_{0j}-2{K_{\alpha j}})\phi_j +(N-2)\beta h\right] +\\
& \sum\limits_{\begin{array}{cc}
{\tiny\tiny{\alpha ,\beta  = 1}}\\
{\tiny\tiny{\alpha  < \beta}}
\end{array}}^N \int_{ - \infty }^{ + \infty }  \ldots  \int_{ - \infty }^{ + \infty } \prod\limits_{k = 1}^N d{\phi _k}\exp \left[ { - \frac{1}{2}\sum\limits_{ij} {{\phi _i}{K_{ij}}{\phi _j}} } +\sum\limits_j (K_{0j}-2{K_{\alpha j}}-2{K_{\beta j}})\phi_j+(N-4)\beta h\right] +\\
&\ldots \\
&\left. +\int_{ - \infty }^{ + \infty }  \ldots  \int_{ - \infty }^{ + \infty } \prod\limits_{k = 1}^N d{\phi _k}\exp \left[ { - \frac{1}{2}\sum\limits_{ij} {{\phi _i}{K_{ij}}{\phi _j}} } +\sum\limits_j -K_{0j}\phi_j+(-N)\beta h\right] \right\} \end{array}
 \end{equation}

Then the partition function now is, 

\begin{equation}\label{eq21}
\begin{array}{ll}
{\rm Q} = & \exp \left( {\frac{1}{2}{\bf{K}}_0{{\bf{K}}^{ - 1}}{{\bf{K}}_0^T}} +N\beta h\right) +
\sum\limits_{\alpha  = 1}^N \exp \left[ {\frac{1}{2}({\bf{K}}_0-2{\bf{K_\alpha}}){{\bf{K}}^{ - 1}}{({\bf{K}}_0-2{\bf{K_\alpha}})^T}}+(N-2)\beta h \right]+\\
& \sum\limits_{\begin{array}{cc}
{\tiny\alpha ,\beta  = 1}\\
{\alpha  < \beta }
\end{array}}^N \exp \left[ {\frac{1}{2}({\bf{K}}_0-2{\bf{K_\alpha}}-2{\bf{K_\beta}}){{\bf{K}}^{ - 1}}{({\bf{K}}_0-2{\bf{K_\alpha}}-2{\bf{K_\beta}})^T}}+(N-4)\beta h \right]+\\
& \sum\limits_{\begin{array}{cc}
{\alpha ,\beta ,\gamma  = 1}\\
{\alpha  < \beta  < \gamma }
\end{array}}^N \exp \left[ {\frac{1}{2}({\bf{K}}_0-2{\bf{K_\alpha}}-2{\bf{K_\beta}}-2{\bf{K_\gamma}}){{\bf{K}}^{ - 1}}{({\bf{K}}_0-2{\bf{K_\alpha}}-2{\bf{K_\beta}}-2{\bf{K_\gamma}})^T}}+(N-6)\beta h \right]+
\ldots \\
& + \exp \left[ {\frac{1}{2}(-{\bf{K}}_0){{\bf{K}}^{ - 1}}{(-{\bf{K}}_0)^T}}+(-N)\beta h \right]
 \end{array}
\end{equation}

\section{Some calculations and discussions}

\subsection{Comparisons of  the partition function with the previous exact results}
\begin{figure}[htb]
 \includegraphics[scale=0.5]{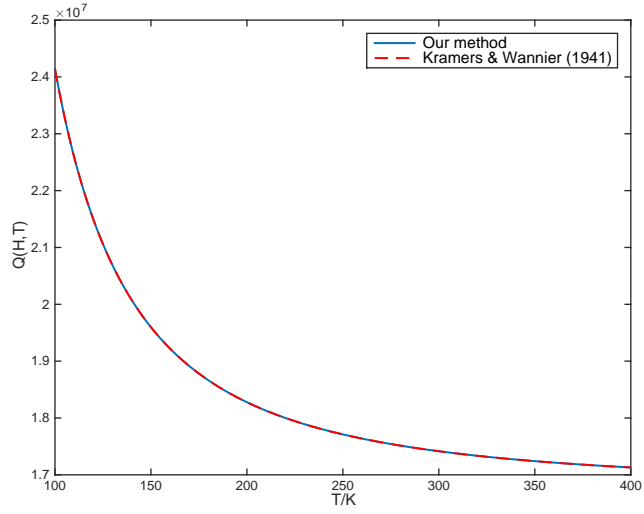}
 \caption{\footnotesize Comparison of the partition function calculate from Eq. (\ref{eq21}) with that from Eq. (\ref{KW}). $N=24$, $\epsilon=1.0\times10^{-3}\mathrm{\  eV}$, $\mu H=1.0\times10^{-3}\mathrm{\  eV}$.}
\label{fig1}
\end{figure}

\begin{figure}[htb]
 \includegraphics[scale=0.5]{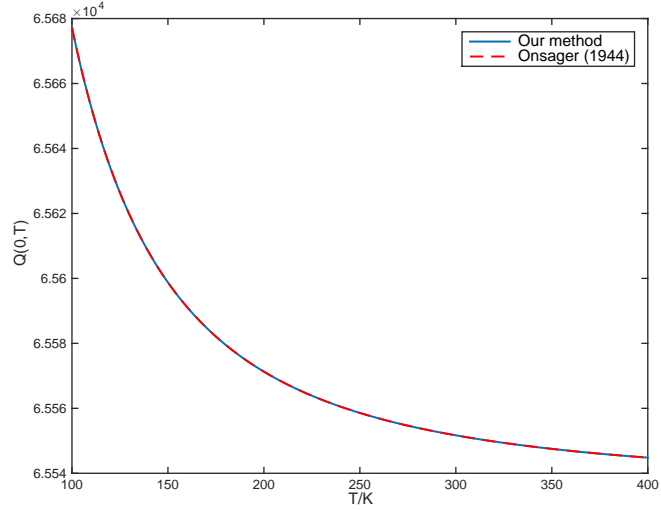}
 \caption{\footnotesize Comparison of the partition function calculate from Eq. (\ref{eq11}) with that from Eq. (\ref{Onsager}). $N=4\times 4$, $\epsilon=1.0\times10^{-4}\mathrm{\ eV}$.}
\label{fig2}
\end{figure}

We compare the partition functions calculated from Eq. (\ref{eq11}) and (\ref{eq21}) with those from Eq. (\ref{KW}) and (\ref{Onsager}), respectively.  The parameters related are: (1) For 1D Ising model, $N=24$, $\epsilon=1.0\times10^{-3}\mathrm{\ eV}$, and $\mu H=1.0\times10^{-3}\mathrm{\ eV}$; (2) For 2D Ising model, $N=4\times 4$ and $\epsilon=1.0\times10^{-4}\mathrm{\  eV}$. (3) Periodic boundary condition is imposed. The results are shown in Fig. \ref{fig1} and Fig. \ref{fig2}, respectively. It can be seen that both partition functions from Eq. (\ref{eq11}) and (\ref{eq21}) are the same with previous exact results (Eq. (\ref{KW}) and (\ref{Onsager}), respectively), which illustrate the correctness of the Eq. (\ref{eq11}) and (\ref{eq21}).  It should be pointed out that the parameters here are selected randomly without special meaning (the same below).  

We can also compare qualitatively the partition functions from Eq. (\ref{eq11}) with that of  Eq. (\ref{KW}) and (\ref{Onsager}) for the case of vanishing field. If Eq. (\ref{KW}) is expanded,  one can get,

\begin{equation}\label{KW_e}
Q(0,T)=2\left\{\exp(N\beta\epsilon)+\mathrm{C}_{_\mathrm{N}}^2\exp[(N-4)\beta\epsilon]+\mathrm{C}_{_\mathrm{N}}^4\exp[(N-8)\beta\epsilon]+...\right\}
\end{equation}

And if Eq. (\ref{Onsager}) is expanded partially, one can get,

\begin{equation}\label{Onsager_e}
\begin{array}{ll}
Q(0,T)&=\left\{\exp(2N\beta\epsilon)+\mathrm{C}_{_\mathrm{N}}^1\exp[(2N-4)\beta\epsilon]+\mathrm{C}_{_\mathrm{N}}^2\exp[(2N-8)\beta\epsilon]+...\exp(-2N\beta\epsilon)\right\}\\
&\times\exp\left[\frac{N}{2\pi}\int_0^\pi {\rm{d}}\phi\log\frac{1}{2}(1+\sqrt{1-\kappa^2\sin^2\phi})\right]
\end{array}
\end{equation}

It is obvious from Eq. (\ref{KW_e}) and (\ref{Onsager_e}) that Eq. (\ref{KW}) or (\ref{Onsager}) can also be expressed by the sum of a series of exponential functions, through which we infer that Eq. (\ref{eq11}) are consistent with them in the form.

\subsection{Partition function of 2D and 3D Ising model}

\begin{figure}[htb]
 \includegraphics[scale=0.5]{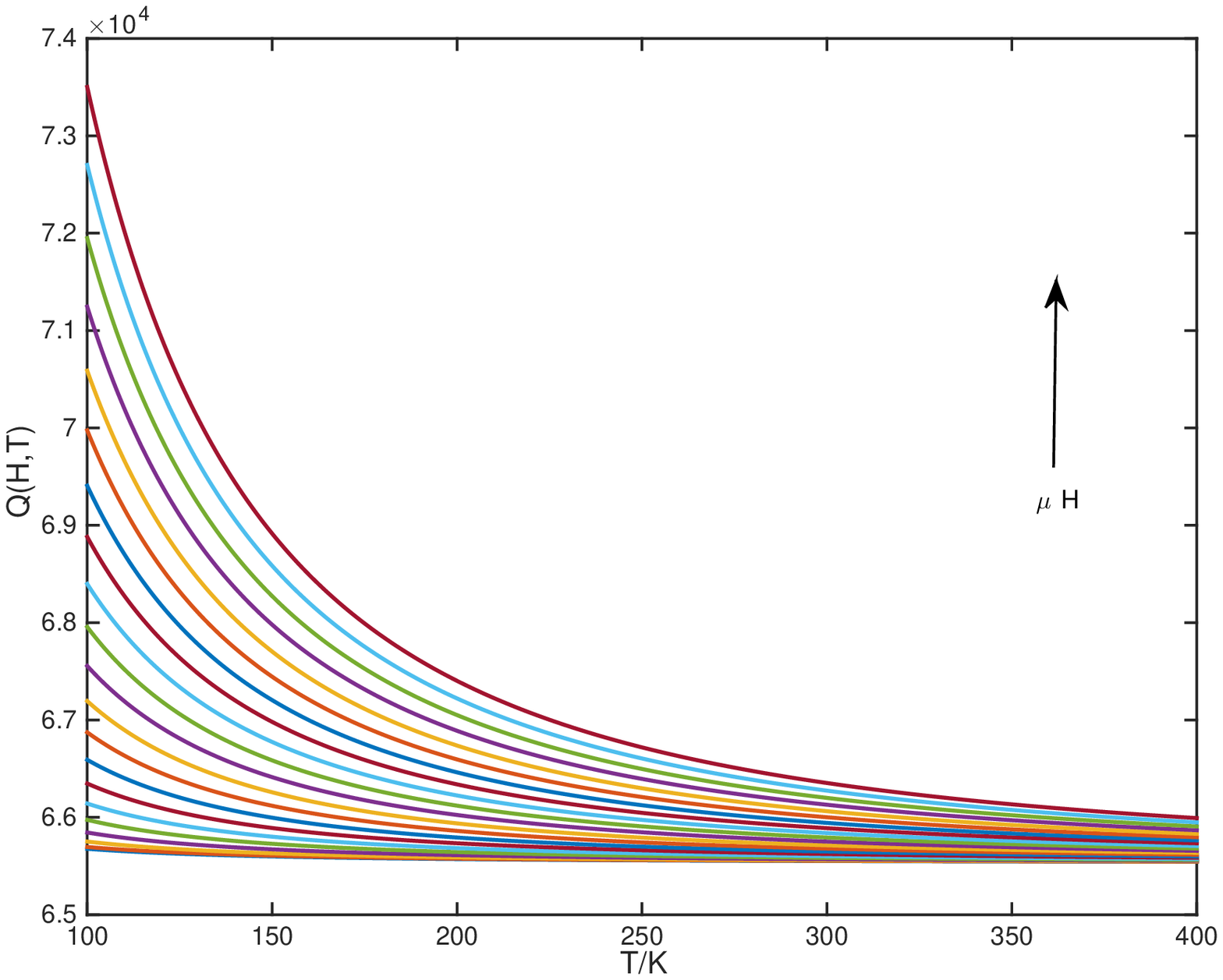}
 \caption{\footnotesize Partition function for 2D Ising model from Eq. (\ref{eq21}). $N=4\times 4$, $\epsilon=1.0\times10^{-4}\mathrm{\ eV}$, $\mu H=0.0-1.0\times10^{-3}\mathrm{\ eV}$ with an interval of $5.0\times 10^{-5}\mathrm{\ eV}$; $T=100.0-400.0\mathrm{\ K}$.}
\label{fig3}
\end{figure}

\begin{figure}[htb]
 \includegraphics[scale=0.5]{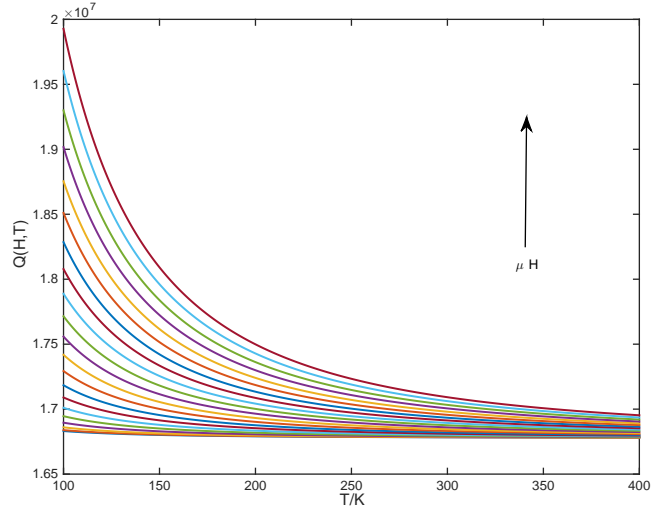}
 \caption{\footnotesize Partition function for 3D Ising model from Eq. (\ref{eq21}). $N=2\times 6\times 2$, $\epsilon=1.0\times10^{-4}\mathrm{\ eV}$, $\mu H=0.0-1.0\times10^{-3}\mathrm{\ eV}$ with an interval of $5.0\times 10^{-5}\mathrm{\ eV}$; $T=100.0-400.0\mathrm{\ K}$.}
\label{fig4}
\end{figure}

We calculate the partition functions for 2D and 3D Ising model from Eq. (\ref{eq21}).  The parameters related are: (1) For 2D case, $N=4\times 4$, $\epsilon=1.0\times10^{-4}\mathrm{\ eV}$, and $\mu H=0.0-1.0\times10^{-3}\mathrm{\ eV}$ with an interval of $5.0\times 10^{-5} \mathrm{\ eV}$; $T=100.0-400.0\mathrm{\ K}$ ; (2) For 3D case, $N=2\times 6\times 2$, and other parameters are the same as the above; (3) Periodic boundary condition is imposed. Results are shown in Fig. \ref{fig3} and Fig. \ref{fig4}, respectively.

It can be seen that the partition functions of both 2D and 3D Ising model increase with the increasing $\mu H$. However, the partition functions decrease with the increasing $T$, which can also be found from Fig. \ref{fig2} when $\mu H=0$ for the 2D Ising model. 

\subsection{Specific heat $C(0,T)$ of a simple 3D Ising model}

With the partition function of Eq. (\ref{eq21}), we can obtain the specific heat, 

\begin{equation}\label{C_C}
C=\frac{\partial E}{\partial T}
\end{equation}

where $$E=-\frac{1}{N}\frac{\partial \log Q}{\partial \beta}$$

\begin{figure}[htb]
 \includegraphics[scale=0.5]{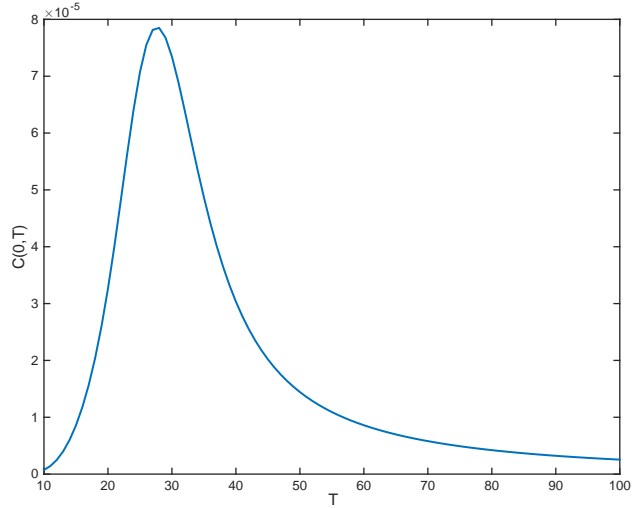}
 \caption{\footnotesize Specific heat $C(0,T)$ of a simple 3D Ising model from Eq. (\ref{C_C}). $N=2\times 6\times 2$, $\epsilon=1.0\times10^{-3}\mathrm{\ eV}$, $\mu H=0.0 \mathrm{\ eV}$; $T=10.0-100.0\mathrm{\ K}$.}
\label{fig5}
\end{figure}

\begin{figure}[htb]
 \includegraphics[scale=0.5]{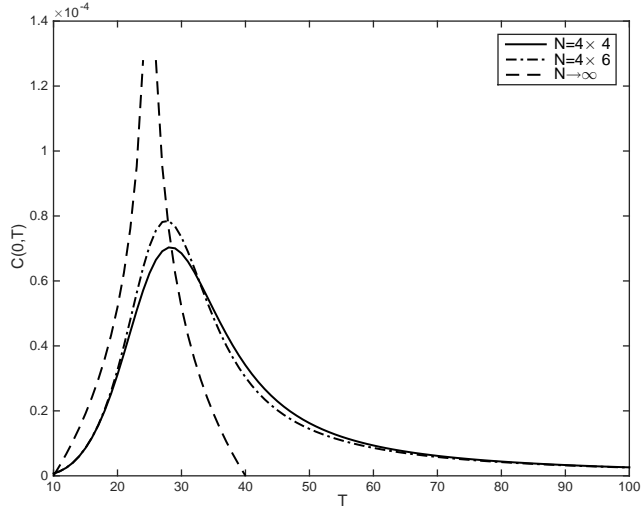}
 \caption{\footnotesize Specific heat $C(0,T)$ of the 2D Ising model for different $N$. $\epsilon=1.0\times10^{-3}\mathrm{\ eV}$, $\mu H=0.0 \mathrm{\ eV}$; $T=10.0-100.0\mathrm{\ K}$.}
\label{fig6}
\end{figure}

Fig. \ref{fig5} shows the specific heat $C(0,T)$ of a simple 3D Ising model varies with the temperature. This model has only $N=2\times 6\times 2$ lattice sites ($\epsilon=1.0\times10^{-3}\mathrm{\ eV}$; $T=10.0-100.0\mathrm{\ K}$). It can be found that specific heat has a maximum at about 30 K but decays rapidly in a short temperature range. This is consistent with experiments or numerical results (eg., Kim et al., 2002; Karandashev et al., 2017). With the increasing $N$, we conjecture that: (1) The maximum of specific heat will increase, and specific heat has a singularity of some kind when $N\rightarrow\infty$ ; (2) The way by which the specific heat approaches its maximum will change like those shown in Fig. \ref{fig6}. It should be pointed out that the example in Fig. \ref{fig6} is based on 2D Ising model, because there is only the results for 1D and 2D Ising model when $N\rightarrow\infty$.  Some more numerical evidences, although also for 2D Ising model, can be also found in Karandashev et al. (2017).

\subsection{Magnetisation $M(T)$ of a simple 3D Ising model}

With the partition function of Eq. (\ref{eq21}), we can obtain the magnetisation $M(T)$, 

\begin{equation}\label{M_ising}
M=\frac{\partial }{\partial (\beta h)} \frac{1}{N} \log Q
\end{equation}

\begin{figure}[htb]
 \includegraphics[scale=0.5]{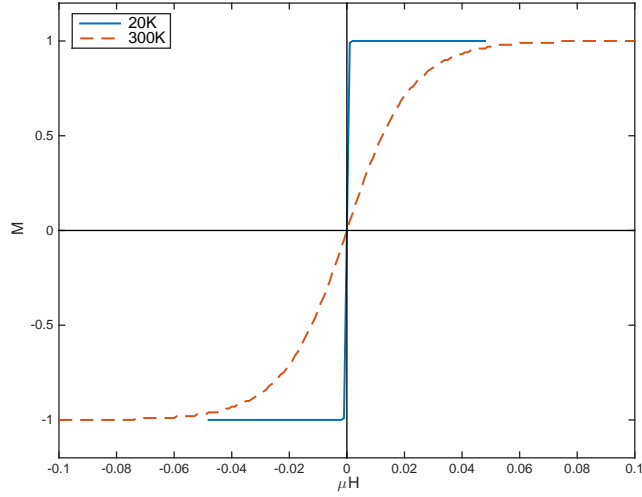}
 \caption{\footnotesize Magnetisation of a simple 3D Ising model varies with the external magnetic field. $\epsilon=1.0\times10^{-3}\mathrm{\ eV}$, $h=-0.1-0.1 \mathrm{\ eV}$.}
\label{fig7}
\end{figure}

Fig. \ref{fig7} shows the $M(T)$ of a simple 3D Ising model varies with an external magnetic field $h=\mu H$. This model has only $N=2\times 6\times 2$ lattice sites ($\epsilon=1.0\times10^{-3}\mathrm{\ eV}$; $h=-0.1-0.1 \mathrm{\ eV}$). The $M(T)$ curves at the temperature of $20\mathrm{\ K}$ and $300\mathrm{\ K}$ are calculated from Eq. (\ref{M_ising}), respectively. It can be found that magnetisation at the low temperatures has different behaviors from those at the high temperatures. This is consistent with experiments or previous results (eg., Kim et al., 2002).

\begin{figure}[htb]
 \includegraphics[scale=0.65]{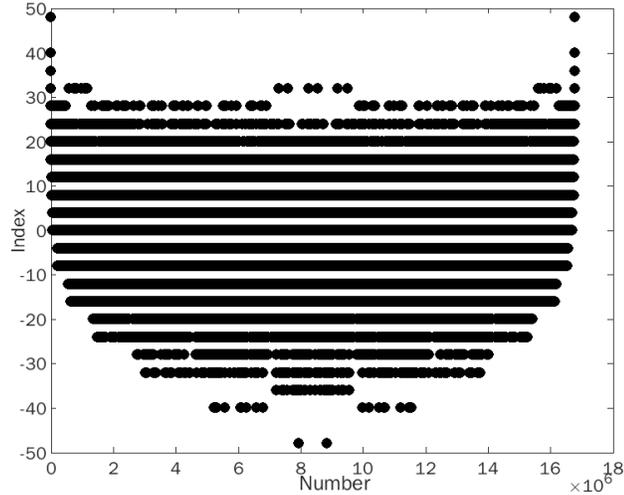}
 \caption{\footnotesize Indexes ($\left({\frac{1}{2}{\bf{K}}_0^T{{\bf{K}}^{ - 1}}{{\bf{K}}_0}} \right)$ etc. when $\epsilon=1$ and $\beta=1$) for a simple 3D Ising model of $N=2\times 6 \times 2$ with periodic boundary.}
\label{fig8}
\end{figure}

\subsection{Some problems in the yielding the partition function}

For an Ising model of $N$ lattice sites, it can be seen from Eq. (\ref{eq11}) or (\ref{eq21}) that we will calculate the sum of $2^N$ exponential functions. For example, if $N=2\times 6 \times 2$, 16777216 operations will be performed. Furthermore, this is not the real difficulty. The real difficulty lies in calculating all possible combinations of the elements of a vector $V=[1\  2\  3 \  ... \  N]$ taken $k, k=1,2,3,....N$ at a time, in order to yield the index of $\left({\frac{1}{2}{\bf{K}}_0{{\bf{K}}^{ - 1}}{{\bf{K}}_0^T}} \right)$, $\frac{1}{2}({\bf{K}}_0-2{\bf{K_\alpha}}){{\bf{K}}^{ - 1}}{({\bf{K}}_0-2{\bf{K_\alpha}})^T}$, ${\frac{1}{2}({\bf{K}}_0-2{\bf{K_\alpha}}-2{\bf{K_\beta}}-2{\bf{K_\gamma}}){{\bf{K}}^{ - 1}}{({\bf{K}}_0-2{\bf{K_\alpha}}-2{\bf{K_\beta}}-2{\bf{K_\gamma}})^T}}$..., when $\epsilon=1$ and $\beta=1$.  Although the algorithm is simple, for a large $N$ the combinations are so large that the finial result might not be exact.   

Fortunately, it seems as if the indexes ($\left({\frac{1}{2}{\bf{K}}_0{{\bf{K}}^{ - 1}}{{\bf{K}}_0^T}} \right)$, $\frac{1}{2}({\bf{K}}_0-2{\bf{K_\alpha}}){{\bf{K}}^{ - 1}}{({\bf{K}}_0-2{\bf{K_\alpha}})^T}$, ${\frac{1}{2}({\bf{K}}_0-2{\bf{K_\alpha}}-2{\bf{K_\beta}}-2{\bf{K_\gamma}}){{\bf{K}}^{ - 1}}{({\bf{K}}_0-2{\bf{K_\alpha}}-2{\bf{K_\beta}}-2{\bf{K_\gamma}})^T}}$... when $\epsilon=1$ and $\beta=1$) distribute according to some rules. Fig. \ref{fig8} shows the distribution of the all 16777216 of these indexes for a simple 3D Ising model of $N=2\times 6 \times 2$ with periodic boundary. It can be found these indexes distribute symmetrically, and most of the indexes are repeated. For this simple 3D Ising model of $N=2\times 6 \times 2$, only 23 indexes are needed to be calculated. This gives us great hope and possibility to calculate these indexes at low cost, at least in principle. However, it is still a difficult problem in this paper to find out the multiplicity of the indexes.

Therefore, for a larger $N$, it is still here an urgent problem how to calculate the sum of $2^N$ exponential functions, especially the Ising model with non-zero external field.

\section{Conclusions}

(1) By using the Hubbard-Stratonovich transformation to express the partition function of the Ising model with $N$ lattice sites in terms of an $N$-dimensional integral, we obtain an exact solution for the partition function of $D$-dimensional Ising model ($D=1,2,3,...$). This solution is expressed in an exponential sum, and does not involve any of complex transcendental function functions or special functions. The most complex calculation is the inverse of a matrix. This solution holds whether or not there is a non-zero external field. 

(2) Although our solution is not elegant as those of Kramers and Wannier (1941) and Onsager (1944) for 1D and 2D Ising model respectively, and involves a large amount of calculation, it at least shows that the Ising model is solvable rigorously in mathematics. The Kramers \& Wannier's solution or the Onsager's solution are consistent with ours in the mathematical form. 

(3) Results from a simple 3D Ising model of $N=2\times 6 \times 2 $ show the specific heat and magnetisation calculated are consistent with those from the experiment observation and numerical simulations.

\vspace{5em}


\ \ 

\end{document}